\documentclass[a4paper,fleqn,usenatbib]{mnras}

\usepackage{txfonts}

\usepackage[T1]{fontenc}
\usepackage{ae,aecompl}

\usepackage{xspace} 
\usepackage{ulem}

\usepackage{graphicx}	

\newcommand {\xmm} {\textit{XMM-Newton}\xspace}
\newcommand {\chandra} {\textit{Chandra}\xspace}
\newcommand {\swift} {\textit{Swift}\xspace}
\newcommand {\h} {H.E.S.S.\xspace}

\newcommand {\src} {HESS\,J1832$-$093\xspace}

\newcommand {\radecposchandra} {$\rm RA=18^h 32^m 45.1^s \pm 0.5'', \rm Dec=-9^\circ 21' 54'' \pm 0.5''~(J2000)$\xspace}

\newcommand {\xmmu}{XMMU\,J183245$-$0921539\xspace}
\newcommand {\irmass}{2MASS\,J18324516$-$0921545\xspace}

\title[\src: A new gamma-ray binary?]{Discovery of a variable X-ray counterpart to \src: A new gamma-ray binary?}

\author[P. Eger et al.]{
P.~Eger$^{1}$\thanks{peter.eger@mpi-hd.mpg.de}, 
H.~Laffon$^{2}$, 
P.~Bordas$^{1}$, 
E.~de~O\~na~Whilhelmi$^{3}$, 
J.~Hinton$^{1}$, 
G.~P\"uhlhofer$^{4}$
\\
$^{1}$Max-Planck-Institut f\"ur Kernphysik, PO Box 103980, 69029 Heidelberg, Germany\\
$^{2}$Universit\'e Bordeaux 1, CNRS/IN2P3, Centre d'\'Etudes Nucl\'eaires
de Bordeaux Gradignan, 33175 Gradignan, France\\
$^{3}$Institute of Space Sciences (CSIC-IEEC), E-08193 Barcelona, Spain\\
$^{4}$Institut f\"ur Astronomie und Astrophysik, Universit\"at  T\"ubingen,
Sand 1, 72076 T\"ubingen, Germany
}

\date{Accepted 2016 January 12; Received 2016 January 12; in original form 2015 November 18}

\pubyear{2016}

\begin{document}
\label{firstpage}
\pagerange{\pageref{firstpage}--\pageref{lastpage}}
\maketitle

\begin{abstract}
The TeV $\gamma$-ray point source \src remains unidentified despite extensive multi-wavelength studies. 
The $\gamma$-ray emission could originate in a very compact 
pulsar wind nebula or an X-ray binary system composed of the X-ray source \xmmu and a companion star (\irmass).
To unveil the nature of \xmmu and its relation to \src, we performed deeper follow-up observations in X-rays with \chandra and \swift to improve source localisation and to investigate time variability.
We observed an increase of the X-ray flux by a factor $\sim$6 in the \chandra data compared to previous observations. 
The source is point-like for \chandra and its updated position is only 0.3$''$ offset from \irmass , confirming the association with this infrared source. Subsequent \swift ToO observations resulted in a lower flux, again compatible with the one previously measured with \xmm, indicating a variability timescale of the order of two months or shorter.
The now established association of \xmmu and \irmass and the observed variability in X-rays are strong evidence for binary nature of \src . 
Further observations to characterise the optical counterpart as well as to search for orbital periodicity are needed to confirm this scenario. 
\end{abstract}




\section{Introduction}\label{sec-introduction}
Extensive observations of the Galactic plane with the High Energy Stereoscopic System (H.E.S.S.) have resulted in the detection of a new population of 
faint $\gamma$-ray sources at TeV energies.
However, the lack of statistics and/or multi-wavelength (MWL) counterparts often prevent a firm identification of those sources. This is the case of \src, a TeV point source close to the Galactic plane recently discovered by H.E.S.S. \citep{HESSJ1832}. 
The TeV spectrum is well described by a powerlaw model with a photon index of 
$\Gamma = 2.6\pm 0.3_\mathrm{stat}\pm 0.1_\mathrm{sys}$ and an integrated photon flux of $I(E>1\,\mathrm{TeV})=(3.0\pm 0.8_\mathrm{stat}\pm 0.6_\mathrm{syst})\times 10^{-13}$\,cm$^{-2}$s$^{-1}$ \citep{HESSJ1832}.
Nearly all Galactic sources seen with \h are extended beyond the instrument's point spread function (PSF) and appear to be mostly related to supernova remnants (SNRs) or pulsar wind nebulae (PWNe). 
Galactic TeV point sources are rather rare and are usually associated with gamma-ray binaries, but can also stem from young compact PWNe, as is the case for G0.9$+$0.1 \citep{Aharonian_G0.9}, or 
even from a background Active Galactic Nucleus (AGN) such as HESS\,J1943$+$213 \citep{hessj1943}.
Only five systems are firmly identified as TeV gamma-ray binaries so far: PSR\,B1259$-$63, LS\,5039, LSI$+$61 303, HESS\,J0632+057 and 1FGL\,J1018.6-5856 
\citep{Aharonian_1259,Aharonian_LS5039,Albert2009,HESSJ0632,J1018}.
It is interesting to note that the latter two systems were first discovered in GeV and/or TeV gamma-rays and were only confirmed later at lower energies through deep follow-up observations. 
All these systems are high-mass X-ray binaries (HMXBs), consisting of a compact object orbiting a massive companion O or Be type star.
Moreover, these sources generally exhibit a peak in their broad-band spectral energy distribution (SED) at MeV-GeV energies, except for HESS\,J0632+057 
which is not detected in this energy band \citep{2013MNRAS.436..740C}.

A search for MWL counterparts was performed to identify the origin of \src. This region was observed in 2011 with \xmm and a hard, highly absorbed X-ray point source was discovered (\xmmu), coincident in position with the TeV emission \citep{HESSJ1832}. 
Furthermore, the infrared (IR) source \irmass ($m_\mathrm{J} = 15.5$, $m_\mathrm{H}=13.2$, $m_\mathrm{K}=12.2$) is located 1.9$^{\prime\prime}$ offset from the \xmm position, compatible within statistical uncertainties, indicating a potential association. The chance probability for such a coincidence is estimated to be $\approx $\,2\% \citep{HESSJ1832}. 
A power-law model fit to the X-ray spectrum of \xmmu resulted in a photon index of $\Gamma=1.3^{+0.5}_{-0.4}$, a column density of $N_H = 10.5 ^{+3.1}_{-2.7}\times 10^{22}\rm~cm^{-2}$ and an unabsorbed energy flux of $\Phi(2-10 \rm~keV) = 6.9 ^{+1.7}_{-2.8}\times 10^{-13} \rm~erg~cm^{-2}~s^{-1}$. 
Such a spectral shape suggested a pulsar nature for \xmmu despite the lack of detected pulsations. Another possibility is that the X-ray source could
originate from the unresolved compact core of an extended PWN. 
However, in the absence of an extended X-ray nebula and/or detected pulsations from \xmmu this scenario could not be confirmed.

On the other hand, the presence of the IR point source only 1.9$''$ offset from \xmmu could indicate that the compact object resides in a binary system with a  stellar companion.
Another argument in favor of this scenario could be the high column density measured for \xmmu which is about a factor 10 larger than the total 
Galactic value seen in HI \citep[$1.5\times10^{22}\rm~cm^{-2}$][]{Dickey1990} in the direction of the source.
Such large column densities may be present in HMXBs due to local material from the dense stellar companion wind or from Be circumstellar discs, as observed for several HMXBs in the Galaxy \citep{SuzakuHMXBs} and the Small Magellanic Cloud \citep{EgerSMCHMXBs}. 
However, the absence of an optical counterpart to \irmass, at least down to magnitudes of $\sim$18, as evidenced by the lack of entries in the USNO 2.0 catalog \citep{USNO} within 20$''$, could also indicate strong interstellar extinction/absorption along this particular line of sight. 
This may be an alternative explanation for the strong absorption seen from \xmmu . 
Furthermore, given the low level of count statistics available, no evidence for variability could be found in the \h or \xmm data which would support the binary scenario.

Alternatively, the X-ray/TeV source could be associated with an AGN located behind the Galactic plane. 
Here the large column density could arise from material in the host galaxy or in the local environment of the AGN. 

To determine the nature of \xmmu, we observed this source again with \chandra in July 2015. 
Its superior angular resolution allows us to probe for smaller source extensions to search for the presence of an X-ray nebula which would confirm the PWN scenario. 
Also, \textit{Chandra's} unprecedented accuracy in absolute astrometry allows us to further investigate the potential association of \xmmu with the 2MASS counterpart. 
The new \chandra results are described in Section~\ref{sec-chandra} and strengthen a gamma-ray binary interpretation for \src.
Following up on these results, we asked for two \swift target of opportunity (ToO) observations in September 2015, which are presented in Section~\ref{sec-swift}. 
Finally we discuss the results obtained in the context of the binary scenario in Section \ref{discussion}.

\section{\chandra observation}\label{sec-chandra}
We observed \xmmu with \chandra\ on July 6, 2015, for 18.2\,ks with the ACIS-I \citep{2003SPIE.4851...28G} camera (ObsID: 16737, P.I.: H.~Laffon). 
To improve the timing resolution to 0.5\,s (down from 3.0\,s in regular mode) the ACIS-I chips were operated in the 1/8 subarray configuration. 
For the analysis of the \chandra data we used the CIAO software version 4.7, supported by tools from the FTOOLS package \citep{1995ASPC...77..367B} and XSPEC version 12.8.1g for spectral modeling \citep{1996ASPC..101...17A}. 
We reprocessed the \texttt{event1} data with the latest calibration version (CALDB 4.6.9) using \texttt{chandra\_repro}. 
The background level in this dataset is very low and the standard good-time-intervals (GTIs) provided by the analysis chain encompass the full exposure time.

\begin{figure}
\resizebox{0.98\hsize}{!}{\includegraphics[]{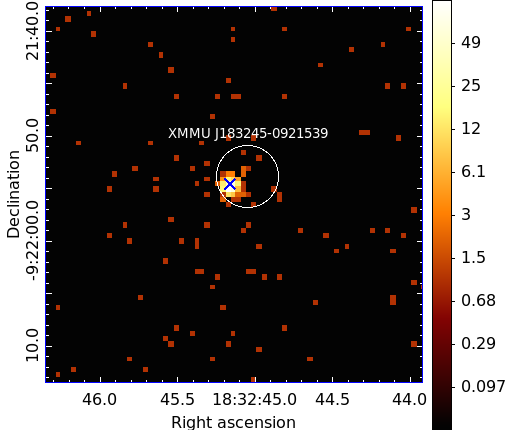}}
\centering
\caption{Counts map of the \chandra observation (0.3--10\,keV). The position and uncertainty (1\,$\sigma$) of \xmmu from the previous \xmm observation is shown as a circle (white). 
The position of the IR source \irmass is represented by the cross (blue).}
\label{fig-chandra-map}
\end{figure}

Figure~\ref{fig-chandra-map} shows the \chandra counts map (0.3--10.0\,keV), zoomed in on the region of \xmmu . 
The source is clearly detected within the positional uncertainty of the original \xmm detection. 
The source detection method \texttt{celldetect} results in a position of \radecposchandra. 
This location is now compatible with the infrared counterpart \irmass within 0.3$^{\prime\prime}$, with a chance coincidence probability of only 0.3\%, and hence confirms this identification. 

\begin{figure}
\resizebox{0.98\hsize}{!}{\includegraphics[]{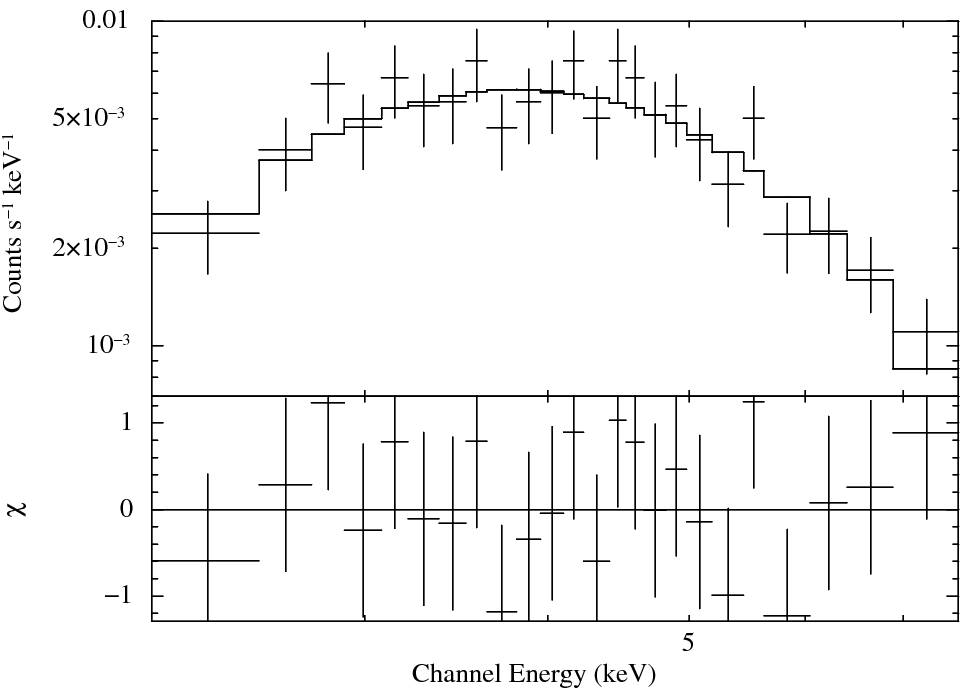}}
\centering
\caption{\chandra spectrum of \xmmu (datapoints with error bars) with the best-fit absorbed powerlaw model (stepped line).}
\label{fig-chandra-spectrum}
\end{figure}

\begin{table*}
\caption[]{Spectral fit results}
\begin{center}
\begin{tabular}{llllll}
\hline\hline\noalign{\smallskip}
\multicolumn{1}{l}{Instrument} &
\multicolumn{1}{l}{MJD} &
\multicolumn{1}{l}{counts} &
\multicolumn{1}{l}{$N_\mathrm{H}$} &
\multicolumn{1}{l}{$\Gamma$} &
\multicolumn{1}{l}{$F_\mathrm{X}$(2-10\,keV)$^{(*)}$} \\
\multicolumn{1}{l}{} &
\multicolumn{1}{l}{(days)} &
\multicolumn{1}{l}{} &
\multicolumn{1}{l}{($10^{22}$\,cm$^{-2}$)} &
\multicolumn{1}{l}{} &
\multicolumn{1}{l}{($10^{-12}$\,erg\,cm$^{-2}$s$^{-1}$)} \\
\noalign{\smallskip}\hline\noalign{\smallskip}
\swift & 54524 & 15$\pm$3.5 & $=10$ & $=1.4$ & 1.0$^{+0.6}_{-0.6}$ \\[0.2cm]
\xmm$^{(**)}$ & 55633 & -- & 10.5$^{+3.1}_{-2.7}$ & 1.3$^{+0.5}_{-0.4}$ & 0.69$^{+0.17}_{-0.28}$ \\[0.2cm]
\swift & 55876 & 2$\pm$1.4 & $=10$ & $=1.4$ & $<2.6$ \\[0.2cm]
\chandra & 57209 & 416$\pm$20 & 9.5$^{+5.2}_{-4.6}$ & 1.5$^{+0.8}_{-0.7}$ & 4.2$^{+1.0}_{-0.6}$ \\[0.2cm]
\swift & 57285 & 26$\pm$6 & $=10$ & $=1.4$ & 1.4$^{+0.6}_{-0.6}$ \\[0.2cm]
\swift & 57291 & 24$\pm$5 & $=10$ & $=1.4$ & 1.2$^{+0.5}_{-0.5}$ \\[0.2cm]
\noalign{\smallskip}\hline
\end{tabular}
\end{center}
\label{tab-fit-results}
The fit results above are for an absorbed powerlaw model. 
Parameters kept fixed in the fit are marked with the '$=$' sign. \\
$^{(*)}$Unabsorbed integrated energy flux.\\
$^{(**)}$Results taken from \citet{HESSJ1832}; number of counts not available, data not analysed in this work.
\end{table*}

\subsection{Spectral analysis}\label{sec-chandra-spectrum}
For the spectral analysis we extracted the source counts from a circular region, centered on the \texttt{celldetect} position, with a radius of 4$^{\prime\prime}$, and the background counts from a nearby source-free region.
 To extract the spectra and response files we used the CIAO tool \texttt{specextract} with a point-like source assumption. 
We fit the spectrum using \texttt{XSPEC} with an absorbed powerlaw model. 
To account for the interstellar photo-electric absorption we used the \texttt{tbabs} model along with the Galactic metal abundances from \citet{wilms2000}. 
The spectrum with the best-fit model is shown in Fig.~\ref{fig-chandra-spectrum}, and the fit results are listed in Tab.~\ref{tab-fit-results}. 
The powerlaw model provides a very good fit ($\chi^2/\mathrm{dof} = 12.75/21$). 
The column density and photon index are compatible within statistical uncertainties with the previous \xmm measurement \citep{HESSJ1832}. 
However, the flux is a factor of 6 higher, indicating significant variability compared to the earlier observation in 2011. 
The spectrum is similarly well described by an absorbed thermal plasma model (MEKAL, $\chi^2/\mathrm{dof} = 11.71/21$). 
However, the best-fit temperature is unrealistically large: $\mathrm{kT} = 15^{+14}_{-9}$\,keV. 

Due to the very limited field of view of the 1/8 subarray configuration and the dithering motion of the \chandra spacecraft, the source position was located outside the active detector area for a significant fraction of the exposure time.
Therefore, the total number of 416 counts in the source region is about a factor of 4 lower than one would expect for an 18\,ks ACIS-I observation, given the flux. This effect is automatically taken into account in the GTIs and we note this here only to avoid confusion when comparing the derived flux with the total number of counts for a 18\,ks \chandra observation.

\subsection{Source morphology}\label{sec-chandra-morphology}
To investigate the source morphology of \xmmu in detail we simulated the telescope point-spread-function (PSF) using the \chandra raytracing simulator \texttt{ChaRT} \citep{chart}. 
As inputs we provided the exact detector position and photon energy distribution of \xmmu as seen with \chandra. 
We then reprojected the simulated PSF onto the detector-plane with the tool \texttt{MARX}. 
Using the CIAO tool \texttt{srcextent} we compared the counts image to the simulated PSF image and determined that the source is not significantly extended for \chandra.
The derived upper limit for the intrinsic source size is 0.28$^{\prime\prime}$ (90\% confidence).

\subsection{Timing analysis}\label{sec-chandra-timing}
\begin{figure}
\resizebox{0.98\hsize}{!}{\includegraphics[]{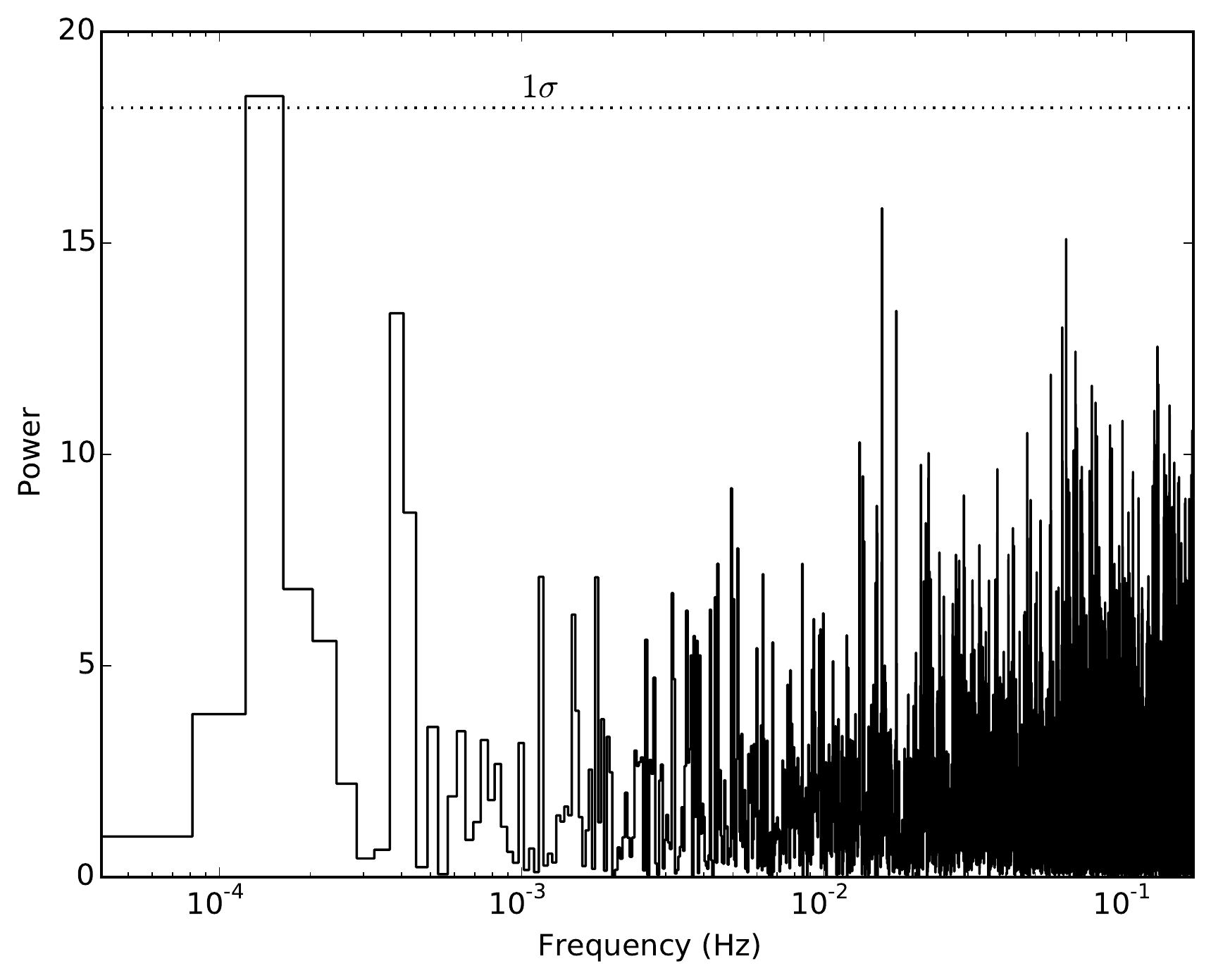}}
\centering
\caption{FFT power spectrum of the \chandra light curve of \xmmu . The 1\,$\sigma$ confidence limit derived from simulations is drawn as a horizontal line (dotted).}
\label{fig-powspec}
\end{figure}

The detection of pulsations from \xmmu would proof its nature as a pulsar and provide a wealth of information regarding the origin of the TeV and X-ray emission. 
For the timing analysis we extracted the 0.3--10\,keV light curve from the same region that was used for the spectral analysis (see Sect.~\ref{sec-chandra-spectrum}) with a binning of 0.5\,s, corresponding to the time resolution of the ACIS-I detector in the 1/8 subarray configuration. 
To search for pulsed emission we computed the Fast-Fourier-Transform (FFT) of the light curve using the FTOOL \texttt{powspec} (see Fig.~\ref{fig-powspec}). 
For the evaluation of the significance of the peaks in power we simulated 10000 light curves of steady sources with the same number of counts and live time, taking also the gaps in exposure due to the \chandra dithering motion into account. 
From the distribution of the maximum power reached in these simulated light curves we derived the 1,2, and 3\,$\sigma$ confidence levels for the detection of pulsations at powers of 18.2, 22.7, and 25.9, respectively. 
As can be seen in Fig.~\ref{fig-powspec}, no significant pulsations are detected at any frequency accessible with the current \chandra dataset. 

To estimate the limit on the pulsed fraction that can be derived from the new \chandra data, we simulated light curves consisting of a flat, unpulsed and a sinusoidal component, again with the same number of counts and time exposure characteristics as in the \chandra data.
We varied the ratio between the number of counts in the flat and periodic components between 0\% and 100\% (with 5\% increments), using pulse periods of 1\,s, 10\,s, 100\,s, and 1000\,s. 
We compared the mean power at the simulated frequency in the FFT power spectra from 1000 simulated light curves for each parameter combination to the 3\,$\sigma$ (99\% confidence) level derived from the simulated unpulsed light curves (see previous paragraph). 
We defined the limit at that pulsed fraction for which the mean power in the simulated light curves falls below the 3\,$\sigma$ level derived from steady sources.
Table~\ref{tab-pulsed-fraction} summarises the upper limits derived in this way for the pulsed fractions at the various pulse periods.

\begin{table}
\caption[]{Limits on the pulsed fraction of \xmmu}
\begin{center}
\begin{tabular}{ll}
\hline\hline\noalign{\smallskip}
\multicolumn{1}{l}{period} &
\multicolumn{1}{l}{pulsed fraction} \\
\multicolumn{1}{l}{(s)} &
\multicolumn{1}{l}{} \\
\noalign{\smallskip}\hline\noalign{\smallskip}
1 & $<$45\%\\
10 & $<$40\%\\
100 & $<$40\%\\
1000 & $<$45\%\\
\noalign{\smallskip}\hline
\end{tabular}
\end{center}
\label{tab-pulsed-fraction}
\end{table}

\begin{figure}
\resizebox{0.98\hsize}{!}{\includegraphics[]{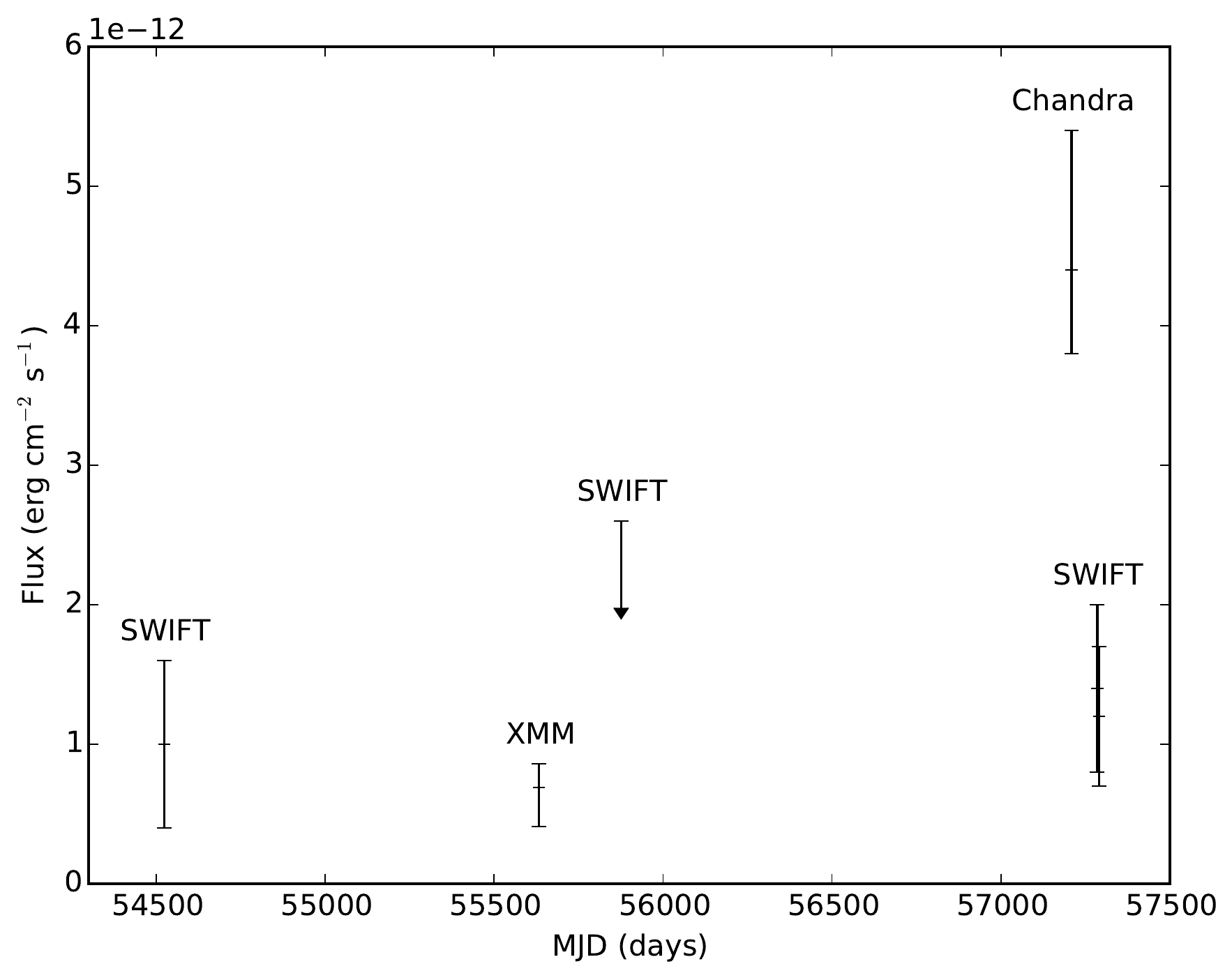}}
\centering
\caption{Long-term light curve of \xmmu including all X-ray observations performed so far. Shown are the unabsorbed integrated energy fluxes (2--10\,keV).}
\label{fig-long-term-lc}
\end{figure}

\section{\textit{Swift}-XRT observations}\label{sec-swift}
\xmmu was serendipitously observed with \swift twice in the past. 
Following up on the flux increase measured with \chandra we asked for two additional dedicated \swift -XRT ToO observations in September 2015, spaced by 6 days. 
All observations are summarised in Tab.~\ref{tab-swift-obs}. 
For the analysis of the \swift -XRT data we used the \texttt{HEASOFT} software package version 6.15 together with the XRT calibration database (\texttt{CALDB}) version 20150721. 
We reprocessed the data using \texttt{xrtpipeline} and provided the resulting \texttt{cleaned} level 2 events (grades 0-12) as input to \texttt{XSELECT} for spectral extraction. 
To extract the source spectra we used a circular region with a radius of 47$^{\prime\prime}$ centered on \xmmu . 
For the background spectra we defined a concentric annular region with inner and outer radii of 94$^{\prime\prime}$ and 188$^{\prime\prime}$, respectively. 
As detector response we used the appropriate file from CALDB (\texttt{swxpc0to12s6\_20010101v014.rmf}) and produced a custom effective area file using \texttt{xrtmkarf}. 

We fitted the spectra with \texttt{XSPEC} assuming an absorbed powerlaw model with all parameters fixed at their best-fit values from the \xmm and \chandra analyses, except for the flux normalisation. 
\xmmu is detected in three \swift -XRT observations, and we calculated an upper limit (99\% confidence) for the fourth, very short (563\,ks) observation, assuming the same spectral shape. 
The results are compiled in Tab.~\ref{tab-fit-results} and we show the evolution of the 2-10\,keV energy flux with time in Fig.~\ref{fig-long-term-lc}. 
All \swift-XRT flux measurements are compatible within uncertainties with the low-state flux seen with \xmm , including the ToO observations that started 76 days after the \chandra detection of \xmmu in the high-flux state. 
Therefore, the source appears to be variable on timescales of $\lesssim$\,2 months or less.

\begin{table}
\caption[]{\swift-XRT observations of \xmmu}
\begin{center}
\begin{tabular}{lll}
\hline\hline\noalign{\smallskip}
\multicolumn{1}{l}{MJD} &
\multicolumn{1}{l}{sequence} &
\multicolumn{1}{l}{exposure} \\
\multicolumn{1}{l}{(days)} &
\multicolumn{1}{l}{no.} &
\multicolumn{1}{l}{(s)} \\
\noalign{\smallskip}\hline\noalign{\smallskip}
54524 & 36174001 & 2409\\
55876 & 44300001 & 563\\
57285$^{(*)}$ & 34056001 & 2610\\
57291$^{(*)}$ & 34056002 & 2837\\
\noalign{\smallskip}\hline
\end{tabular}
\end{center}
\label{tab-swift-obs}
All data in photon counting (PC) mode\\
$^{(*)}$new ToO observations following the \chandra result. 
\end{table}

\section{Discussion}\label{discussion}

We observed an increase of a factor $\sim$6 in the X-ray flux of the source \xmmu during the \chandra observation (see Fig. \ref{fig-long-term-lc}). 
Moreover, the source is point-like for \chandra and can only be extended at the sub-arcsecond level.  
Furthermore, its position is now determined with unprecedented accuracy and is only 0.3$''$ offset from the potential IR counterpart \irmass which could be the  companion star of \xmmu. 
Therefore the new \chandra results provide strong evidence in favor of a binary origin of \src.
Except for the flux, the other spectral parameters (powerlaw index, column density) do not show any significant change compared to previous observations. 
We note, however, that the statistical uncertainties of the column density in the new \chandra data are relatively large, and the non-detection of significant variability in this parameter may be due to a lack of sufficient statistics. 
Such differences are expected along the orbit of a binary system and were observed e.g. for HESS\,J0632$+$057 \citep{Aliu2014}. 
Alternatively, the large column density could also arise from additional interstellar molecular material in the Galactic plane, not traced by HI measurements. 

Also, it is currently not yet excluded that the gamma-ray/X-ray source is associated with an extragalactic object such as an AGN, being also an alternative explanation for the very large column density measured in X-rays. 
In this case, the IR emission would come from dust in the host galaxy and the X-ray and gamma-ray emission would be produced by inverse Compton scattering of accelerated particles in the jet. 
The additional absorption component may arise from interstellar material in the host galaxy or from matter in direct vicinity of the black hole. 
As already discussed by \citet{HESSJ1832}, due to the hard X-ray spectrum such an AGN should be most likely a flat spectrum radio quasar (FSRQ) \citep[see, e.g.][]{1997MNRAS.284..569P}. 
One drawback of this scenario is the fact that one would also expect to see high-energy emission in the GeV band from inverse Compton scattering. 
However, there is no source in the \textit{Fermi}-LAT 3FGL catalog at this position and an upper limit of $3.6 \times 10^{-12} \rm\,erg\,cm^{-2}\,s^{-1}$ in the $10\rm\,GeV-100\rm\,GeV$ energy band was derived in \citet{HESSJ1832}. 
Furthermore, the measured spectral index in the VHE gamma-ray regime \citep[$\Gamma$=2.6,][]{HESSJ1832} would be much harder than observed from any other FSRQ in this energy range, such as PKS\,1510$-$089 \citep{pks1510} and PKS\,1222$+$21 \citep{pks1222}. 
If the origin of the emission from \src is indeed extragalactic, this would be  a very unusual object, belonging to a yet unknown class of VHE gamma-ray emitting AGN. 

\begin{figure}
\resizebox{0.98\hsize}{!}{\includegraphics[]{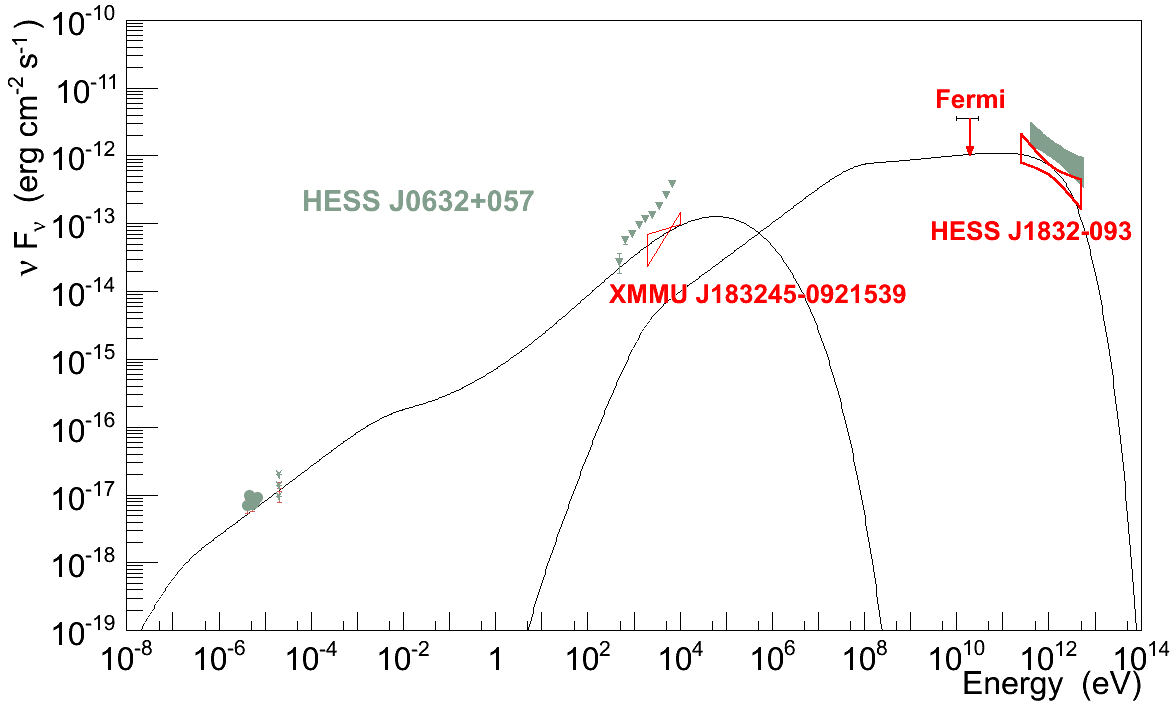}}
\centering
\caption{Comparison of the SEDs of the two sources \src (red) and HESS\,J0632$+$057 (grey). 
The data for \src (\xmm, \textit{Fermi}-LAT upper limit, \h) are from \citet{HESSJ1832}. For HESS\,J0632$+$057 we show the multi-wavelength data (from left to right: GMRT, VLA, \xmm, \h) from \citet{Skilton2009}. 
The solid line (grey) depicts the same model as used for HESS\,J0632$+$057 \citep{Hinton2009,Skilton2009}, only scaled by a factor of $\sim3$ to fit the data of \src .}
\label{fig-sed}
\end{figure}

Since the MWL features of \src do not resemble commonly observed extragalactic sources, we assume in the following that the source originates in a Galactic binary system.
We note that there is only one object firmly identified as a gamma-ray binary which does not exhibit any GeV emission: HESS\,J0632$+$057 \citep{2013MNRAS.436..740C}.
Interestingly, the TeV and X-ray spectra as well as the overall flux level seen from \src are remarkably similar to this TeV gamma-ray binary.
To demonstrate this similarity, we compare the SEDs for both sources in  Fig.~\ref{fig-sed}. The X-ray data shown for both sources are from \xmm observations during which the sources were in a low-flux state. The model curve overplotted is a one-zone time-dependent synchrotron and inverse Compton model, assuming an injection electron spectral index of 2.0, used to describe HESS\,J0632$+$057 by \citet{Skilton2009} and scaled down by a factor $\sim$3 to fit the SED from \src. 
Also, follow-up observations in radio of \src could further test this scenario, as known TeV gamma-ray binaries exhibit hard, non-thermal spectra in this wavelength regime.

Further similarities between \src and HESS\,J0632$+$057 can be tentatively considered in light of the relative fluxes observed in their corresponding X-ray light-curves. Although the sampling for \src is rather sparse, it already provides a flux ratio $\sim$6 between high- and low/quiescent-states, here identified respectively with the \chandra and the \xmm /\swift data points. About the same ratio is indeed observed in the X-ray light-curve of HESS\,J0632$+$057 between peak-to-base-line flux levels, with maximum X-ray fluxes occurring at orbital phases $\sim$0.3 \citep{Bongiorno-2011}. Furthermore, the typical time-scales of rise and decay of the X-ray flaring episodes in HESS\,J0632$+$057 are found to be of the order of a few weeks, which is so far consistent with the X-ray measurements of \src presented here, constraining the variability timescale to $\lesssim$2~months. Interestingly, HESS\,J0632$+$057 displays a double peak structure in the X-ray light-curve (see also the case for other gamma-ray binaries, e.g. LS\,I\,$+$61~303, \citealp{Smith-2009, Li-2011} or 1FGL\,J1018.6$-$5856, \citealp{An-2015}). The second X-ray peak occurs at orbital phases $\in$~[0.6--0.9], being significantly shallower and more variable from orbit to orbit with respect to the main X-ray peak \citep{Aliu2014}. If similarities of \src were to extend also to the presence of such a secondary peak, future X-ray monitoring observations of the source may reveal it, further strengthening the gamma-ray binary scenario. 

The X-ray spectra of \src obtained with the different instruments display a rather hard and relatively stable photon index (within statistical uncertainties) for the different observations (see Tab.~\ref{tab-fit-results}). VHE emission in gamma-ray binaries is usually interpreted as Inverse Compton (IC) radiation of highly relativistic electrons scattering off the companion's photon field. For typical companion temperatures, $T_{\star} \in$~[22500 -- 39000]~K and TeV electrons, IC emission proceeds deep into the Klein-Nishina (KN) regime, with cooling time-scales $t_{\rm KN} \approx 1.7 \times 10^{2} \, w_{100}^{-1} \,E_{\rm e~TeV}^{0.7}$~s, where $w_{100}$ is the companion photon field energy density in units of 100 erg~cm$^{-3}$ and $E_{\rm e~TeV}$ is the electron energy in TeV \citep[see, e.g.,][]{Khangulyan-2008}. The X-ray emission is thought to originate from synchrotron radiation by the same population of relativistic electrons, with synchrotron emission time-scales $t_{\rm sy} \approx 4 \times 10^{2}\, B_{\rm G}^{-2} \, E_{\rm e~TeV}^{-1}$, with $B_{\rm 1G}$ the magnetic field in Gauss units. 

The higher energy output observed in VHE gamma-rays compared to that in X-rays (see SED in Fig.~\ref{fig-sed}) suggests that the IC channel is dominant over synchrotron. Given that $t_{\rm KN} < t_{\rm sy}$ and assuming that IC losses take place deep in the KN regime, it implies relatively hard electron spectra, with electron spectral indices $P_{\rm e} \lesssim 2$, resulting in a hard synchrotron spectrum in X-rays. 

In the case of \src, marginal evidence that similar processes are responsible for the X-ray and TeV emission may come from the spectral properties measured at both energy bands, $\Gamma_{\rm X} \in$~[1.3--1.5] and $\Gamma_{\gamma} = 2.6 \pm 0.4$ \citep{HESSJ1832}. However, a strong caveat here is the non-detection of the optical counterpart of \src, which prevents a more refined characterisation in such a gamma-ray binary scenario. Future observations, either with highly sensitive instruments operating in the optical-UV band and/or high-resolution IR spectrography in case absorption prevents a proper optical characterisation of the companion star, will hopefully provide in-depth information of the physical properties of this new TeV binary candidate.  

\section{Conclusion}\label{conclusion}
The new X-ray observations of \xmmu using \chandra and \swift shed new light on the nature of the gamma-ray source \src.
The PWN scenario, as discussed as one of the possible origins in \citet{HESSJ1832}, is not supported due to the lack of any measured extension of the X-ray source beyond the \chandra resolution. 
The binary scenario seems most likely due to the detected X-ray flux variability and the broadband SED, although an extragalactic origin from a very unusual VHE gamma-ray emitting AGN, located directly in the line of sight of the Galactic plane is not excluded. 
Unfortunately, the currently scarce coverage of observations available in X-rays does not allow a precise measurement of the variability time scale of this object.
Future monitoring of the source in X-rays with \textit{Swift} will hopefully permit the orbital period of the system to be determined, as in the case of HESS\,J0632$+$057 \citep{Aliu2014}.
Deeper observations in IR and radio could also help to identify the nature of the IR source \irmass.

\section*{Acknowledgements}
This research has made use of data obtained from the Chandra Data Archive and software provided by the Chandra X-ray Center (CXC) in the application package CIAO. 
This publication makes use of data products from the Two Micron All Sky Survey, which is a joint project of the University of Massachusetts and the Infrared Processing and Analysis Center/California Institute of Technology, funded by the National Aeronautics and Space Administration and the National Science Foundation. 
We thank V. Zabalza for his useful input to the scientific discussion.




\bibliographystyle{mnras}
\bibliography{citations}

\begin{thebibliography}{}
\makeatletter
\relax
\def\mn@urlcharsother{\let\do\@makeother \do\$\do\&\do\#\do\^\do\_\do\%\do\~}
\def\mn@doi{\begingroup\mn@urlcharsother \@ifnextchar [ {\mn@doi@}
  {\mn@doi@[]}}
\def\mn@doi@[#1]#2{\def\@tempa{#1}\ifx\@tempa\@empty \href
  {http://dx.doi.org/#2} {doi:#2}\else \href {http://dx.doi.org/#2} {#1}\fi
  \endgroup}
\def\mn@eprint#1#2{\mn@eprint@#1:#2::\@nil}
\def\mn@eprint@arXiv#1{\href {http://arxiv.org/abs/#1} {{\tt arXiv:#1}}}
\def\mn@eprint@dblp#1{\href {http://dblp.uni-trier.de/rec/bibtex/#1.xml}
  {dblp:#1}}
\def\mn@eprint@#1:#2:#3:#4\@nil{\def\@tempa {#1}\def\@tempb {#2}\def\@tempc
  {#3}\ifx \@tempc \@empty \let \@tempc \@tempb \let \@tempb \@tempa \fi \ifx
  \@tempb \@empty \def\@tempb {arXiv}\fi \@ifundefined
  {mn@eprint@\@tempb}{\@tempb:\@tempc}{\expandafter \expandafter \csname
  mn@eprint@\@tempb\endcsname \expandafter{\@tempc}}}

\bibitem[\protect\citeauthoryear{{Abramowski} et~al.,}{{Abramowski}
  et~al.}{2011}]{hessj1943}
{Abramowski} A.,  et~al., 2011, \mn@doi [\aap] {10.1051/0004-6361/201116545},
  \href {http://adsabs.harvard.edu/abs/2011A%26A...529A..49H} {529, A49}

\bibitem[\protect\citeauthoryear{{Aharonian} et~al.,}{{Aharonian}
  et~al.}{2005a}]{Aharonian_G0.9}
{Aharonian} F.,  et~al., 2005a, \mn@doi [\aap] {10.1051/0004-6361:200500022},
  \href {http://adsabs.harvard.edu/abs/2005A%26A...432L..25A} {432, L25}

\bibitem[\protect\citeauthoryear{{Aharonian} et~al.,}{{Aharonian}
  et~al.}{2005b}]{Aharonian_1259}
{Aharonian} F.,  et~al., 2005b, \mn@doi [\aap] {10.1051/0004-6361:20052983},
  \href {http://adsabs.harvard.edu/abs/2005A%26A...442....1A} {442, 1}

\bibitem[\protect\citeauthoryear{{Aharonian} et~al.,}{{Aharonian}
  et~al.}{2006}]{Aharonian_LS5039}
{Aharonian} F.,  et~al., 2006, \mn@doi [\aap] {10.1051/0004-6361:20065940},
  \href {http://cdsads.u-strasbg.fr/abs/2006A%26A...460..743A} {460, 743}

\bibitem[\protect\citeauthoryear{{Aharonian} et~al.,}{{Aharonian}
  et~al.}{2007}]{HESSJ0632}
{Aharonian} F.~A.,  et~al., 2007, \mn@doi [\aap] {10.1051/0004-6361:20077299},
  \href {http://adsabs.harvard.edu/abs/2007A%26A...469L...1A} {469, L1}

\bibitem[\protect\citeauthoryear{{Albert} et~al.,}{{Albert}
  et~al.}{2009}]{Albert2009}
{Albert} J.,  et~al., 2009, \mn@doi [\apj] {10.1088/0004-637X/693/1/303}, \href
  {http://cdsads.u-strasbg.fr/abs/2009ApJ...693..303A} {693, 303}

\bibitem[\protect\citeauthoryear{{Aleksi{\'c}} et~al.,}{{Aleksi{\'c}}
  et~al.}{2011}]{pks1222}
{Aleksi{\'c}} J.,  et~al., 2011, \mn@doi [\apjl] {10.1088/2041-8205/730/1/L8},
  \href {http://adsabs.harvard.edu/abs/2011ApJ...730L...8A} {730, L8}

\bibitem[\protect\citeauthoryear{{Aliu} et~al.,}{{Aliu}
  et~al.}{2014}]{Aliu2014}
{Aliu} E.,  et~al., 2014, \mn@doi [\apj] {10.1088/0004-637X/780/2/168}, \href
  {http://adsabs.harvard.edu/abs/2014ApJ...780..168A} {780, 168}

\bibitem[\protect\citeauthoryear{{An} et~al.,}{{An} et~al.}{2015}]{An-2015}
{An} H.,  et~al., 2015, \mn@doi [\apj] {10.1088/0004-637X/806/2/166}, \href
  {http://adsabs.harvard.edu/abs/2015ApJ...806..166A} {806, 166}

\bibitem[\protect\citeauthoryear{{Arnaud}}{{Arnaud}}{1996}]{1996ASPC..101...17A}
{Arnaud} K.~A.,  1996, in {G.~H.~Jacoby \& J.~Barnes} ed.,  Astronomical
  Society of the Pacific Conference Series Vol. 101, Astronomical Data Analysis
  Software and Systems V. p.~17

\bibitem[\protect\citeauthoryear{Blackburn}{Blackburn}{1995}]{1995ASPC...77..367B}
Blackburn J.~K.,  1995, Astronomical Data Analysis Software and Systems IV, 77,
  367

\bibitem[\protect\citeauthoryear{{Bongiorno}, {Falcone}, {Stroh}, {Holder},
  {Skilton}, {Hinton}, {Gehrels}  \& {Grube}}{{Bongiorno}
  et~al.}{2011}]{Bongiorno-2011}
{Bongiorno} S.~D.,  {Falcone} A.~D.,  {Stroh} M.,  {Holder} J.,  {Skilton}
  J.~L.,  {Hinton} J.~A.,  {Gehrels} N.,   {Grube} J.,  2011, \mn@doi [\apjl]
  {10.1088/2041-8205/737/1/L11}, \href
  {http://adsabs.harvard.edu/abs/2011ApJ...737L..11B} {737, L11}

\bibitem[\protect\citeauthoryear{{Caliandro} et~al.,}{{Caliandro}
  et~al.}{2013}]{2013MNRAS.436..740C}
{Caliandro} G.~A.,  et~al., 2013, \mn@doi [\mnras] {10.1093/mnras/stt1615},
  \href {http://adsabs.harvard.edu/abs/2013MNRAS.436..740C} {436, 740}

\bibitem[\protect\citeauthoryear{{Carter}, {Karovska}, {Jerius}, {Glotfelty}
  \& {Beikman}}{{Carter} et~al.}{2003}]{chart}
{Carter} C.,  {Karovska} M.,  {Jerius} D.,  {Glotfelty} K.,   {Beikman} S.,
  2003, in {Payne} H.~E.,  {Jedrzejewski} R.~I.,   {Hook} R.~N.,  eds,
  Astronomical Society of the Pacific Conference Series Vol. 295, Astronomical
  Data Analysis Software and Systems XII. p.~477

\bibitem[\protect\citeauthoryear{{Dickey} \& {Lockman}}{{Dickey} \&
  {Lockman}}{1990}]{Dickey1990}
{Dickey} J.~M.,  {Lockman} F.~J.,  1990, \mn@doi [\araa]
  {10.1146/annurev.aa.28.090190.001243}, \href
  {http://adsabs.harvard.edu/abs/1990ARA%26A..28..215D} {28, 215}

\bibitem[\protect\citeauthoryear{{Eger} \& {Haberl}}{{Eger} \&
  {Haberl}}{2008}]{EgerSMCHMXBs}
{Eger} P.,  {Haberl} F.,  2008, \mn@doi [\aap] {10.1051/0004-6361:200810809},
  \href {http://adsabs.harvard.edu/abs/2008A%26A...491..841E} {491, 841}

\bibitem[\protect\citeauthoryear{Garmire, Bautz, Ford, Nousek  \&
  Ricker~Jr}{Garmire et~al.}{2003}]{2003SPIE.4851...28G}
Garmire G.~P.,  Bautz M.~W.,  Ford P.~G.,  Nousek J.~A.,   Ricker~Jr G.~R.,
  2003, in Society of Photo-Optical Instrumentation Engineers (SPIE) Conference
  Series.

\bibitem[\protect\citeauthoryear{{H.E.S.S.~Collaboration}
  et~al.,}{{H.E.S.S.~Collaboration} et~al.}{2013}]{pks1510}
{H.E.S.S.~Collaboration} et~al., 2013, \mn@doi [\aap]
  {10.1051/0004-6361/201321135}, \href
  {http://adsabs.harvard.edu/abs/2013A%26A...554A.107H} {554, A107}

\bibitem[\protect\citeauthoryear{{H.E.S.S. Collaboration} et~al.,}{{H.E.S.S.
  Collaboration} et~al.}{2015a}]{HESSJ1832}
{H.E.S.S. Collaboration} et~al., 2015a, \mn@doi [\mnras]
  {10.1093/mnras/stu2148}, \href
  {http://adsabs.harvard.edu/abs/2015MNRAS.446.1163H} {446, 1163}

\bibitem[\protect\citeauthoryear{{H.E.S.S. Collaboration} et~al.,}{{H.E.S.S.
  Collaboration} et~al.}{2015b}]{J1018}
{H.E.S.S. Collaboration} et~al., 2015b, \mn@doi [\aap]
  {10.1051/0004-6361/201525699}, \href
  {http://adsabs.harvard.edu/abs/2015A%26A...577A.131H} {577, A131}

\bibitem[\protect\citeauthoryear{{Hinton} et~al.,}{{Hinton}
  et~al.}{2009}]{Hinton2009}
{Hinton} J.~A.,  et~al., 2009, \mn@doi [\apjl] {10.1088/0004-637X/690/2/L101},
  \href {http://adsabs.harvard.edu/abs/2009ApJ...690L.101H} {690, L101}

\bibitem[\protect\citeauthoryear{{Khangulyan}, {Aharonian}  \&
  {Bosch-Ramon}}{{Khangulyan} et~al.}{2008}]{Khangulyan-2008}
{Khangulyan} D.,  {Aharonian} F.,   {Bosch-Ramon} V.,  2008, \mn@doi [\mnras]
  {10.1111/j.1365-2966.2007.12572.x}, \href
  {http://adsabs.harvard.edu/abs/2008MNRAS.383..467K} {383, 467}

\bibitem[\protect\citeauthoryear{{Li} et~al.,}{{Li} et~al.}{2011}]{Li-2011}
{Li} J.,  et~al., 2011, \mn@doi [\apj] {10.1088/0004-637X/733/2/89}, \href
  {http://adsabs.harvard.edu/abs/2011ApJ...733...89L} {733, 89}

\bibitem[\protect\citeauthoryear{{Monet}}{{Monet}}{1998}]{USNO}
{Monet} D.~G.,  1998, in American Astronomical Society Meeting Abstracts. p.
  120.03

\bibitem[\protect\citeauthoryear{{Morris}, {Smith}, {Markwardt}, {Mushotzky},
  {Tueller}, {Kallman}  \& {Dhuga}}{{Morris} et~al.}{2009}]{SuzakuHMXBs}
{Morris} D.~C.,  {Smith} R.~K.,  {Markwardt} C.~B.,  {Mushotzky} R.~F.,
  {Tueller} J.,  {Kallman} T.~R.,   {Dhuga} K.~S.,  2009, \mn@doi [\apj]
  {10.1088/0004-637X/699/1/892}, \href
  {http://adsabs.harvard.edu/abs/2009ApJ...699..892M} {699, 892}

\bibitem[\protect\citeauthoryear{{Padovani}, {Giommi}  \& {Fiore}}{{Padovani}
  et~al.}{1997}]{1997MNRAS.284..569P}
{Padovani} P.,  {Giommi} P.,   {Fiore} F.,  1997, \mnras, \href
  {http://adsabs.harvard.edu/abs/1997MNRAS.284..569P} {284, 569}

\bibitem[\protect\citeauthoryear{{Skilton} et~al.,}{{Skilton}
  et~al.}{2009}]{Skilton2009}
{Skilton} J.~L.,  et~al., 2009, \mn@doi [\mnras]
  {10.1111/j.1365-2966.2009.15272.x}, \href
  {http://adsabs.harvard.edu/abs/2009MNRAS.399..317S} {399, 317}

\bibitem[\protect\citeauthoryear{{Smith}, {Kaaret}, {Holder}, {Falcone},
  {Maier}, {Pandel}  \& {Stroh}}{{Smith} et~al.}{2009}]{Smith-2009}
{Smith} A.,  {Kaaret} P.,  {Holder} J.,  {Falcone} A.,  {Maier} G.,  {Pandel}
  D.,   {Stroh} M.,  2009, \mn@doi [\apj] {10.1088/0004-637X/693/2/1621}, \href
  {http://adsabs.harvard.edu/abs/2009ApJ...693.1621S} {693, 1621}

\bibitem[\protect\citeauthoryear{{Wilms}, {Allen}  \& {McCray}}{{Wilms}
  et~al.}{2000}]{wilms2000}
{Wilms} J.,  {Allen} A.,   {McCray} R.,  2000, \mn@doi [\apj] {10.1086/317016},
  \href {http://adsabs.harvard.edu/abs/2000ApJ...542..914W} {542, 914}

\makeatother
\end{thebibliography}

\bsp	
\label{lastpage}
\end{document}